\documentclass[11pt,twoside]{article}

\usepackage{asp2004}
\usepackage{epsf}
\usepackage{graphicx}

\begin{document}
\title{GMASS, Unveiling Distant Massive Galaxies with Spitzer}  
\author{J.D. Kurk}   
\affil{MPIA, K\"onigstuhl 17, 69117, Heidelberg, Germany\\ INAF -- OAA, Largo E. Fermi 5, 50125, Florence, Italy}
\author{A. Cimatti}   
\affil{INAF -- OAA, Largo E. Fermi 5, 50125, Florence, Italy}
\author{E. Daddi, M. Dickinson}   
\affil{NOAO - Tucson, 950 North Cherry Avenue, Tucson, AZ 85719, USA}
\author{M. Mignoli, L. Pozzetti, M. Bolzonella, S. Berta, G. Zamorani}
\affil{INAF -- OAB, Via Ranzani 1, 40127, Bologna, Italy}
\author{P. Cassata, G. Rodighiero, A. Franceschini}
\affil{Universit\`a di Padova, Vicolo dell'Osservatorio 5, 35122, Padova, Italy}
\author{A. Renzini}
\affil{INAF -- OAP, Vicolo dell'Osservatorio 5, 35122, Padova, Italy}
\author{P. Rosati}   
\affil{ESO, Karl-Schwarzschild-Strasse 2, 85748, Garching, Germany}

\begin{abstract} 
  One of the main questions of galaxy formation is how and when galaxies
  assembled their mass.  Observations suggest that most mass assembly and star
  formation occured at $1 < z < 3$.  However, in the same redshift range,
  fully assembled old and massive galaxies co-exist with star-bursting systems
  still in their assembling stage.  In order to explore galaxy evolution in
  this critical redshift range, we started a project called GMASS (Galaxy Mass
  Assembly ultra-deep Spectroscopic Survey) based on an 145 hours ESO Large
  Program and aimed at doing ultradeep spectroscopy with the VLT of a sample
  of high redshift galaxies slected with IRAC/Spitzer (m$_{\rm AB}(4.5\mu$m$)
  < 23, z_{\rm phot} > 1.4$) from the GOODS-South/HUDF region.  The selection
  at 4.5$\mu$m is crucial to select high redshift galaxies based on the
  stellar mass or objects affected by strong dust extinction.  The
  spectroscopic deep integration times (15 -- 50 hours) are needed to identify
  unambiguously the redshift and nature of these galaxies, which redshifts are
  mostly in the \emph{desert} where spectroscopy is traditionally difficult
  due to the absence of strong emission lines at optical wavelengths.  The
  survey is almost complete and some preliminary results are presented here.
\end{abstract}

\keywords{Galaxies: distances and redshifts --
          Galaxies: evolution --
          Galaxies: formation --
          Galaxies: fundamental parameters --
          Galaxies: high-redshift }


\section{Introduction}   

To find out how the galaxies in the local universe have evolved and formed,
more distant galaxies have been sought and studied in detail.  As the yellow
light of older stars, responsible for most of the stellar mass, shifts out of
the optical and NIR bands, only the younger, short-lived component of the
stellar population can be measured with earth-based telescopes and it becomes
difficult to estimate how much mass is contained within the stars of such
distant galaxies.  To find distant galaxies, techniques have been applied
which use rest-frame UV light to separate distant from nearby systems
\citep[e.g.,][]{steidel2003} and are therefore biased towards the strongly
star forming galaxies, while it has become clear recently that the more
massive galaxies (M $> 10^{11}$ M$_\odot$) have probably undergone a main
period of star formation in the very distant past
\citep[e.g.,][]{cimatti2004,daddi2005} and exhibit low star formation activity
already since many Gyrs ago \citep[e.g.,][]{juneau2005}.  Here the mid- and
far-IR capabilities of Spitzer come to the rescue as they allow us to select
distant galaxies independent of their star formation rate and to estimate the
stellar mass content more reliably.  Still, to determine the redshift and
physical properties of these MIR selected galaxies, knowledge of rest-frame
UV/optical absorption (and emission) lines is necessary.  Especially for
galaxies at $1.4 < z < 2.5$ this is not an easy task as the strongest lines
move out of the observed-frame optical wavelength regime, where spectroscopic
capabilities are still the most powerful, resulting in a \emph{redshift
  desert} where few spectroscopically confirmed galaxies are known \citep[but
see also][]{steidel2004}.  The redshift range of $1 < z < 3$, however, seems
to be just the range over which most of the mass in galaxies was built
\citep{poli2003}, as evidenced by the substantial population of old, massive
and passive E/S0 galaxies at $z \sim 1$ with minimum formation redshifts of
$z_f \sim 2$ \citep{dickinson2003a,fontana2003}.  In particular,
\citet{daddi2004} identified spectroscopically a few galaxies at $z \sim 2$
with redder colours and larger stellar masses (M $>$ 10$^{11}$ M$_\odot$) than
Lyman break galaxies \citep[M $\sim 10^{10}$ M$_\odot$,][]{papovich2001}.  The
SFRs, morphologies and clustering of these galaxies suggest that they may be
the progenitors of the massive $z < 1$ ellipticals, but the small sample
prevents a full understanding of the global picture.  For these reasons, we
have carried out an ESO VLT Large Program (P.I. A.~Cimatti), called GMASS (the
Galaxy Mass ultra-deep Spectroscopic Survey), to obtain spectra of a sample of
massive and/or dusty galaxies in the redshift desert.  The integration times
per mask reach up to 32 hours and for some individual galaxies more than 60
hours.

\section{Sample selection}

The instrument of choice for this survey was FORS2 \citep{appenzeller1998},
which provides excellent sensitivity around 8000\,\AA\ and is stable enough to
combine observations carried out over several months.  FORS2 has a field of
view of $6.8\arcmin\times6.8\arcmin$.  The sample was selected within one
$6.8\arcmin\times6.8\arcmin$ region (see Fig.\ 1) in the \emph{Chandra Deep
  Field South} \citep{giacconi2000}, partly overlapping with the \emph{K20
  field} \citep{cimatti2002} and the \emph{Hubble Ultra Deep Field}
\citep{beckwith2003}.  A wealth of multi-wavelength data is available for this
field as part of the GOODS legacy \citep{dickinson2003b}, from radio to X-ray
wavelengths, including deep Spitzer observations.  The GMASS sample catalog
contains all the objects detected in the IRAC 4.5\,$\mu$m image with
magnitude\footnote{All magnitudes are in the AB system, unless otherwise
  noted.}  $m(4.5\,\mu{\rm m}) < 23.02$.  Selection in the 4.5\,$\mu$m band
provides the best compromise between sensitivity, point spread function, image
quality and blending among the four IRAC bands, takes advantage of the
negative K-correction due to the shift of the 1.6\,$\mu$m bump into this band
(for $z > 1.5$ objects), and provides better sensitivity to stellar masses
(for $z > 2$ objects) and less to dust extinction than the $K$ band.  The
magnitude limit was driven by the compromise between photometric completeness
and sample size.  All of these 1278 $m(4.5\,\mu{\rm m}) < 23.02$ objects are
also detected in $K_s$ band, with m($K_s) < 21.0$ (Vega), except for two.
Using the $K_s$ image as detection image, matched photometric catalogs in all
available optical bands (from $U$ to $H$) were created and used for the
determination of photometric redshifts with HyperZ \citep{bolzonella2000}.
The subsample selected for spectroscopic observations was limited to those
objects with $z_{\rm phot} > 1.4$ and no known (or planned) spectroscopic
redshift from the other surveys in the CDFS \citep[amongst
others:][]{vanzella2005, lefevre2004, szokoly2004}.  The total available time
of 145 hours (including overheads) was distributed among six masks: three
using the blue 300V grism (exposure times of 12, 14, and 15 hours) and three
using the red 300I grism (exposure times of 11, 32, and 30 hours) with
resolutions of 440 and 660, respectively.  Targets had to have magnitudes of
$B < 26.5$ to be included in a blue mask or $I < 26.0$ to be included in a red
mask.  A total of 205 objects fulfill all requirements, 169 of these were
included in the masks.  As the objects included in the mask are selected by
geometrical requirements, the observed sample is (relatively) unbiased.
Remaining slit positions were set to include possibly interesting
\emph{filler} objects.

\begin{figure}[!th]
\vspace{0.3cm}
\begin{center}
\includegraphics[]{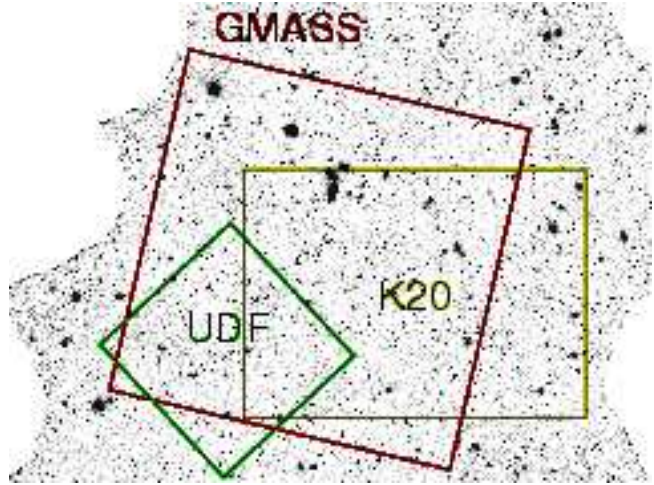}
\end{center}
\caption{Northern part of an ACS/HST F850LP image of the GOODS-South region
  with the GMASS, K20 and UDF fields indicated.}
\end{figure}

\begin{figure}[!bh]
\begin{center}
\includegraphics[width=13.0cm,clip=]{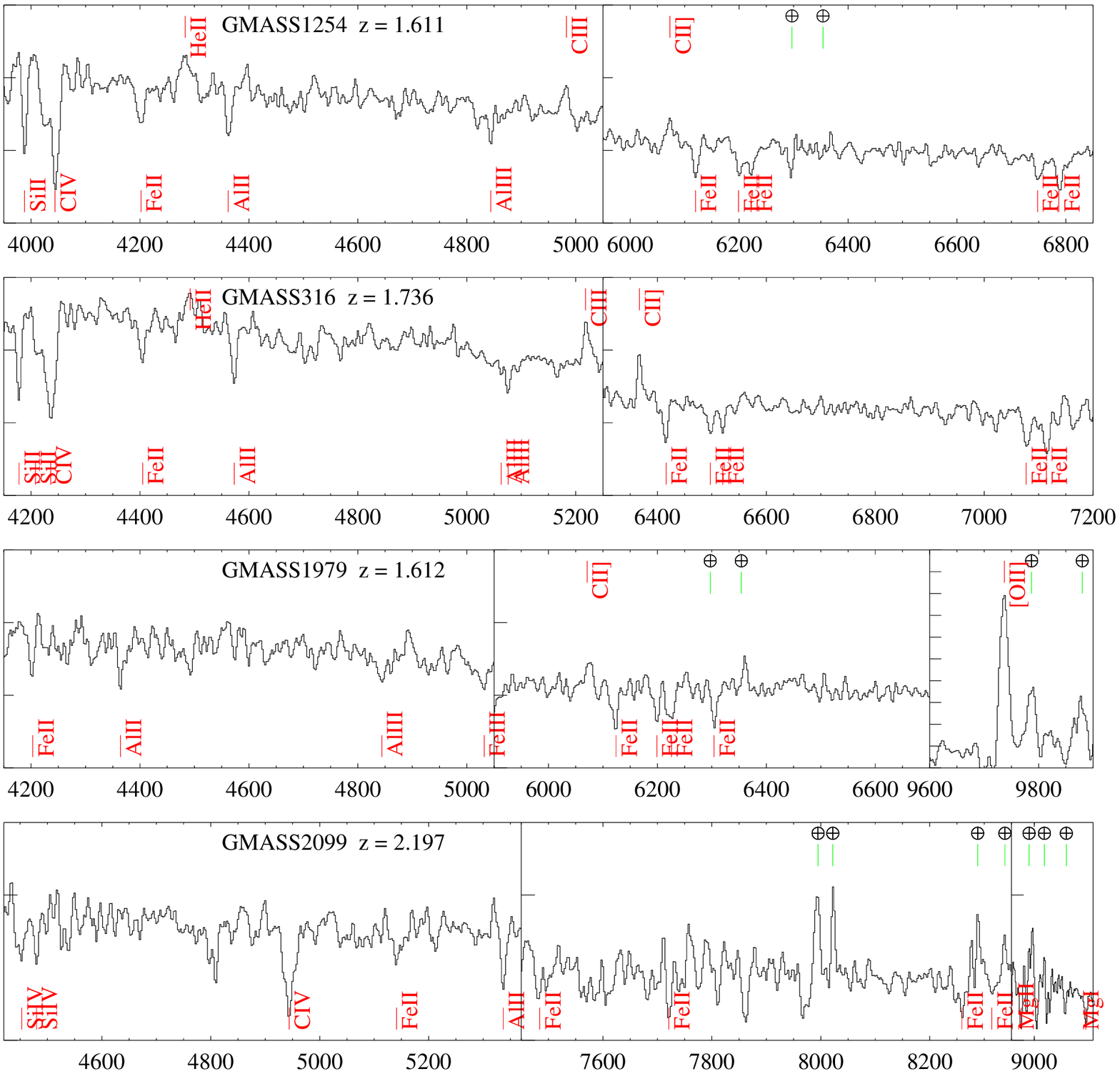}
\end{center}
\caption{Examples of GMASS spectra (from the 12 hour blue mask) zoomed on
  wavelength regions where prominent features are located.  Telluric lines and
  emission lines are indicated above the continuum, while absorption lines are
  indicated below.  The horizontal axis represents observed wavelength in \AA
  ngstr\"om, while the vertical axis is an arbitrary flux scale. Note the
  scale change for GMASS1979 at $\lambda > 9600$\,\AA.}
\end{figure}

\section{Observations and first results}

The observations were carried out in service mode from August 2004 until
December 2005.  Five of the six masks have been reduced and preliminary
redshifts have been determined.  The spectra are of excellent quality; the
brighter targets show plenty of absorption lines (see Fig.\ 2).  As a first
analysis after extracting the spectra, we have attempted to determine
redshifts of the galaxies, currently for the first four masks (all blue masks
and the 11 hour red mask).  Absorption and, if present, emission lines were
fit to the observed features.  In all masks at least a few objects do not show
features strong enough to be fitted.  For these, a redshift has been
determined by cross-correlating with an averaged spectrum of the objects with
known redshifts in the same kind of mask (blue or red).  The cross-correlation
has also been tested on objects for which features could be identified and the
results are consistent.  The total number of redshifts determined (in the
first four masks) is 168, but some of these are for fillers.  The number of
redshifts for the sample of GMASS objects with $z_{\rm phot} > 1.4$ is 102 (a
few of these had reliable spectroscopic redshifts from other surveys).  A
redshift histogram for the GMASS sample is shown in Fig.\ 3.  Shown are the
photometric redshifts for all objects (including those too faint for optical
spectroscopy), the spectroscopic redshifts for those objects for which it is
known (from the literature and from the GMASS project) and the spectroscopic
redshifts determined from GMASS spectra only.  The GMASS project already
contributes 78\% to the number of spectroscopically determined redshifts
greater than 1.4 in this field.

\begin{figure}[!ht]
\plotone{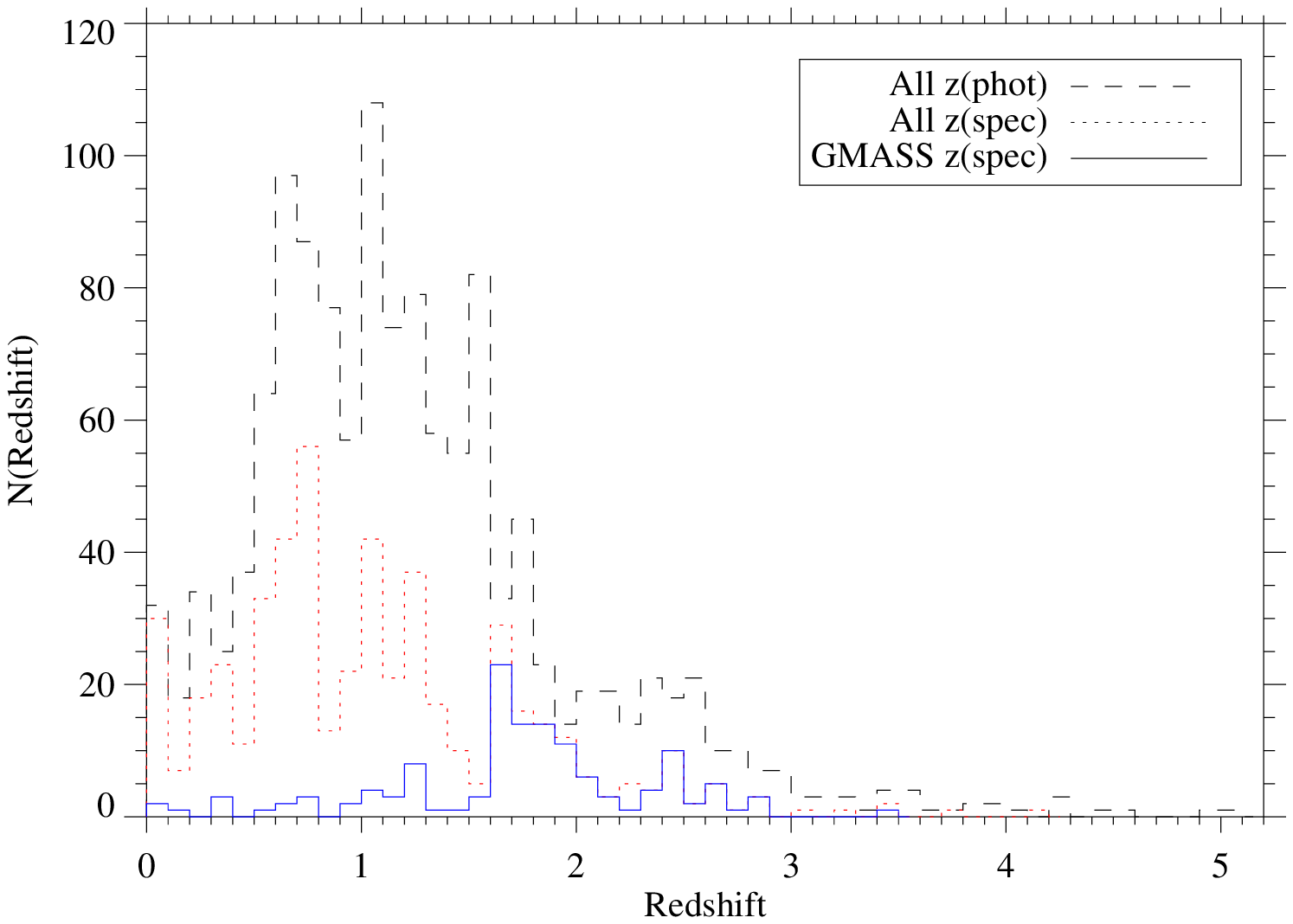}
\caption{Histogram of redshifts in the GMASS field.  The (black) dashed line
  represents the photometric redshifts determined for all 1276 objects.  The
  (red) dotted line represents all the (reliable) spectroscopic redshifts
  known (503), while the (blue) solid line represents the (reliable)
  spectroscopic redshifts determined from the first four GMASS masks (132).}
\end{figure}

\section{Outlook}

With the $\sim 140$ new redshifts $> 1.4$ and over 200 galaxy spectra expected
to be delivered by the GMASS project, great progress in the understanding of
galaxy formation at $1.4 < z < 2.5$ is expected to be made.  We will derive
the evolution of the rest-frame near-IR luminosity function and density to $z
\sim 2 - 3$ to a depth one magnitude deeper than previous studies
\citep{pozzetti2003}.  Using the wide wavelength coverage of the CDFS field,
we are fitting spectral energy distributions (SEDs) up to 8\,$\mu$m, which
will provide estimates of the stellar masses.  Using the whole sample, we will
trace the evolution of the galaxy stellar mass function and thereby
$\Omega_*(z)$ to $z > 2$.  A detailed morphological analysis (e.g., merger
fraction, CAS parameters, surface brightness profiles) of the deep ACS/HST
images of the targets is currently being carried out.  After identification of
different galaxy types from spectroscopy, photometry and morphology, we will
study the evolution of their three-dimensional clustering.  Finally, we will
compare all the obtained results with predictions from current galaxy
evolution models.  Targets which are too faint in the optical but have a red
SED may be followed up with NIR spectroscopy.


\acknowledgements 

We would like to thank the GOODS team for making available the IRAC data of
the CDFS for our purposes.  Based on observations made at the European
Southern Observatory, Paranal, Chile (ESO programme 173.A-0687).  JK
acknowledges financial support from Sonderforschungsbereich 439 \emph{Galaxien
  im jungen Universum} of the Deutsche Forschungsgemeinschaft.


\end{document}